# S-ConvNet: A Shallow Convolutional Neural Network Architecture for Neuromuscular Activity Recognition Using Instantaneous High-Density Surface EMG Images

*Md. Rabiul Islam[1], Daniel Massicotte[1], Francois Nougarou[1], Philippe Massicotte[1] and Wei-Ping Zhu[2]*

[1] Dept. of Electrical and Computer Engineering, Université du Québec à Trois-Rivières, QC, Canada
[2] Dept. of Electrical and Computer Engineering, Concordia University, Montreal, QC, Canada

*Abstract-* The concept of neuromuscular activity recognition using instantaneous high-density surface electromyography (HD-sEMG) images opens up new avenues for the development of more fluid and natural muscle-computer interfaces. However, the existing approaches employed a very large deep convolutional neural network (ConvNet) architecture and complex training schemes for HD-sEMG image recognition, which requires the network architecture to be *pre-trained* on a very large-scale labeled training datasets, as a result it makes computationally very expensive. To overcome this problem, we propose S-ConvNet and All-ConvNet models, a simple yet efficient framework for learning instantaneous HD-sEMG images from scratch for neuromuscular activity recognition. Without using any pre-trained models, our proposed S-ConvNet and All-ConvNet demonstrate very competitive recognition accuracy to the more complex state of the art for neuromuscular activity recognition based on instantaneous HD-sEMG images, while using a $\approx 12 \times$ smaller dataset and reducing learning parameters to a large extent. The experimental results proved that the S-ConvNet and All-ConvNet are highly effective for learning discriminative features for instantaneous HD-sEMG image recognition especially in the data and high-end resource constrained scenarios.

*Index Terms*— Neuromuscular activity recognition, Shallow convolutional neural networks, Feature learning, HD-sEMG, Gesture recognition, Muscle-computer interface, Deep neural networks

## I. Introduction

Neuromuscular activity recognition has been a driving motivation for research because of its respective novel applications in real life. The major application domains are non-invasive control of active prosthesis [1], wheelchairs [2], exoskeletons [3] or providing interaction methods for video games [4] and neuromuscular diagnosis [5]. The conventional approaches for neuromuscular activity recognition immensely rely on multi-channel surface electromyography (sEMG) sensors and windowed descriptive and discriminatory sEMG features [6-10]. However, the sparse multi-channel sEMG based methods are not suitable for real-world applications due to its limitations to electrode shift and positioning and therefore malfunctioning in any one of the channels requires retraining the entire system [11], [12]. To overcome this problem, the high- density sEMG (HD-sEMG) based methods have been proposed in recent years [11-13]. The HD-sEMG records myoelectric signals using two-dimensional (2D) electrode arrays that characterize the spatial distribution of myoelectric activity over the muscles that reside within the electrode pick-up area [14], [15]. The collected HD-sEMG data are spatially correlated which enabled both temporal and spatial changes and robust against malfunction of the channels with respect to the previous counterparts [12]. However, the existing HD-sEMG based neuromuscular activity recognition methods are still depending on the windowed sEMG which demands to find an optimal window length otherwise influence in the classification accuracy and controller delay especially in the application of assistive technology, physical rehabilitation and human computer interfaces [13].

To overcome this problem and develop a more fluid and natural muscle-computer interfaces (MCI's), more recently, Geng *et al.*, [13] and Islam *et al.,* [16] explored the patterns inside the instantaneous sEMG images spatially composed from HD-sEMG enables neuromuscular based gesture recognition solely with the sEMG signals recorded at a specific instant. In their approach, the instantaneous values of HD-sEMG signals at each sampling instant were arranged in a 2D grid in accordance with the electrode positioning. Afterwards, this 2D grid was converted to a grayscale sEMG image. Using Histogram of Oriented Gradients (HOG) as discriminative features and pairwise SVM's classification method in [16], a competitive neuromuscular activity recognition accuracy of an 8-hand gesture has been achieved as par with the state-of-the-art method for an intra-subject test.

However, the state-of-the-art methods [13], [17] employed a DeepFace [18] like very large deep convolutional neural network (ConvNet) architecture for sEMG image classification, which requires to be pre-trained on a very large-scale training dataset, as a result it makes computationally very expensive to be practical for real-world MCI's applications. In their large deep ConvNet includes two locally connected layer (LCN) and three fully connected layers among the other convolutions (CNN) and a $G$-way fully connected layer. The LCN layers are different from the CNN layers in a sense that it assigns an independent filter weight to each of the local receptive field in each layer, while CNN layers adopt a filter weight sharing strategy [19]. More explicitly, in CNN weight sharing strategy, the feature map of a local receptive field is acquired for a given sEMG image $I$ by forming a patch $I_p$ with a shared kernel, $\theta$ i.e. $f_p = I_p^T \theta$, where $p$ stands for the position of the patch in the sEMG image. Therefore, CNN is capable to learn translation/location invariant features by adopting the local connectivity and weight sharing strategies. In contrast, the feature map of a local receptive filed in an LCN layer is acquired by forming a patch with an independent kernel, $\theta_p$ i.e. $f_p = I_p^T \theta_p$. Due to this unshared weight strategies, the LCN layer fails to model the relations of parameters in different locations. Also, the number of learning parameters increases considerably from $m$ to $m \times k$, where $m \gg k$, where $m$ is the number of patches and $k$ is the number of kernels. As a result, a very large-scale labeled training datasets are required to train the LCN layer [18], [19]. However, the LCN layer may be useful in an application where the precise location of the feature is dependent of the class labels with the huge cost of collecting a large-scale labeled dataset ($\approx$ 4.4 million as suggested in [18]) and computational burden of introducing a more sophisticated alignment algorithm to the network structure, though the later part of this claim is ignored and have not been discussed in the existing methods for instantaneous HD-sEMG image recognition.

By considering the above-mentioned fact, we must investigate the most important and a very basic research questions, while designing a new CNN-based architecture for instantaneous HD-sEMG image classification- *(i) Do we expect the devised model to produce a location/translation invariant feature representation?* or, *(ii) do we need a location dependent feature representation?* If the answer is *YES* to the first question, then we can let up the LCN layer while designing a CNN-based architecture for instantaneous HD-sEMG image classification and hence may reduce the burden of gathering a large-scale labeled training dataset significantly and surplus a requirement of introducing a robust alignment algorithm.

Apart from that, the network architecture employed by the existing approaches are heavily rely on pre-trained on large-scale HD-sEMG training datasets ($\approx$ 0.72 millions). The conventional paradigm of using *pre-trained* models in the literature when the *source task A* is different from the *target task B* and when there are not enough *target data* available to make the training accomplishable alone on the *target task B* [20]. However, in the existing approaches for instantaneous HD-sEMG image recognition (e.g., [13], [17]), the *source task A* and the *target task B* are the same, and the *pre-training* on large-scale HD-sEMG training datasets have been performed with the assumption of preventing overfitting while *re-training* with the training data available for the *target task B* i.e intra-subject test. Therefore, it is not surprising to achieve high *target task* accuracy with this highly *resource- based* and *fined-tuned* network architecture. Nevertheless, this longstanding conventional wisdom of pre-training have recently been falsified by He *et al.* [20], where *pre-training* does not necessarily reduce overfitting or improve the final *target task* accuracy is proved to be claimed.

Moreover, there are other critical limitations of using *pre-trained* networks for instantaneous HD-sEMG image classification: *(i) Constrained structure design space-* pre-trained networks are very deep and large and trained on a large-scale HD-sEMG datasets, therefore, containing a massive number of parameters. Hence, there is a little flexibility to control/adjust the network structures (even for a small changes) by directly adopting the *pre-trained* network to the *target task*. The requirement of computing resources and large-scale pre-trained datasets are also bounded by the large network structures. *(ii) Domain mismatch-* sEMG signals are highly subject specific and the distributions of the sEMG signals vary considerably even between recording sessions of the same subject within the same experimental set up [17]. This problem becomes more challenging, where the learned model is used to recognize muscular activities in a new recording session. Though the fine-tuning of the pre-trained model can reduce the gap due to the deformations in a new recording session of the *target task*. However, it's still a serious problem, when there is a huge mismatch between the *source task* and the *target task* [21]. *(iii) Learning bias*- the distributions and the loss functions between the *source task* and the *target task* may vary significantly, which may lead to different searching/optimization spaces. Therefore, the learning may be biased towards a local minimum which is not the optimal for the *target task* [22].

To overcome these above-mentioned problems, our work is motivated by the following research question- *is it possible to train the neuromuscular activity recognition model from the scratch?* To achieve this goal, we propose shallow convolutional neural network (S-ConvNet) architectures, a simple yet effective framework, which could learn neuromuscular activity from scratch using only the makeshift HD-sEMG database available for the *target task*. S-ConvNet is reasonably flexible, which can be tailored to various network structures for different computing platforms such as desktop, server, mobile and even embedded electronics. Though it's being simple and flexible, the S-ConvNet also help keep competitive performance on final *target task* accuracy as par with very complex state of the art methods while reducing the learning parameters to a large extent.

For instantaneous sEMG image based neuromuscular activity recognition, the challenge remains open because very limited research has been done on it. We present S-ConvNet, according to the best of our knowledge, this is the first framework that can train instantaneous HD-sEMG based neuromuscular activity recognition model from scratch with the competitive performance as par with very complex state of the art methods.

Following subsections II and III presents the proposed framework, model description and design principles for S-ConvNet respectively. Section IV presents the All-ConvNet. Section V provides experimental results and discusses the performance of the proposed S-ConvNet and All-ConvNet models for instantaneous HD-sEMG based neuromuscular activity recognition.

## II. THE PROPOSED FRAMEWORK

The proposed framework for neuromuscular activity recognition using instantaneous HD-sEMG images includes following three major computational components: (i) pre-processing and sEMG image generation (ii) architectural design of the S-ConvNet model and (iii) classification. A schematic diagram of the proposed framework of muscular activity recognition by instantaneous sEMG images are shown in Fig. 1. First, the power-line interferences were removed from the acquired HD-sEMG signals with a band-stop filtered between 45 and 55 Hz using a $2^{nd}$ order Butterworth filter. Then, the HD-sEMG signals at each sampling instant were arranged in a 2-D grid according to their electrode positioning. This grid was further transformed into an instantaneous sEMG image by linearly transforming the values of sEMG signals from $mV$ to color intensity as $[-2.5mV, 2.5mV]$ to $[0\ 255]$. Thus, an instantaneous grayscale sEMG image was formed with the size of $16 \times 8$. Secondly, we devised different S-ConvNet models which describes in Section III. Finally, providing instantaneous HD-sEMG images and their corresponding labels, our devised S-ConvNet model is trained offline to predict to which muscular activity an instantaneous HD-sEMG image belongs. Then, this trained S-ConvNet model is used to recognize different neuromuscular activities online from the instantaneous HD-sEMG images.

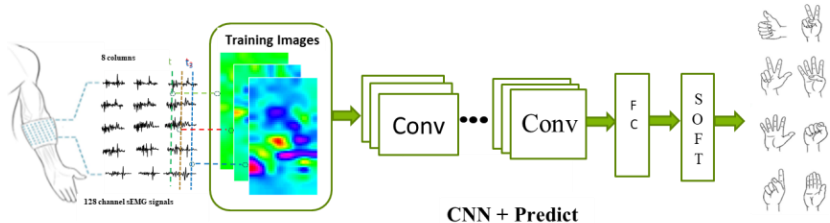

Fig. 1. Schematic diagram of the proposed framework of muscular activity recognition by instantaneous sEMG images

## III. MODEL DESCRIPTION- THE SHALLOW CONVOLUTIONAL NEURAL NETWORK (S-CONVNET)

The proposed S-ConvNet network architectures differs from existing approaches for HD-sEMG image recognition in several key aspects. Firstly, our S-ConvNet models are trained from *random initialization i.e. from scratch without any pre-training*. The pre-training in the existing approaches (e.g., [13], [17]) involves over .72 million of images acquired from 18 different sulobjects. However, considering the targeted application domains of sEMG based neuromuscular activity recognition (e.g. assistive technology, physical rehabilitation etc.), it is always difficult to gather such a large amount dataset required for the pre-training of a very deep neural network models. We can't expect an amputee or a patient to provide a large set of training examples over multiple number of trials and sessions. Moreover, there is no evidence if this specialized very large and deep neural network architectures are required for models to be pre-trained for instantaneous HD-sEMG image recognition. Our work demonstrates that it is often possible to match the accuracy of highly *resource- based* and *fined-tuned* network architecture when training from scratch even using a simple S-ConvNet network architectures. Training from *random initialization*, our S-ConvNet models requires ($\approx$ 12 × *smaller dataset* than its pre-trained counterparts for HD-sEMG image recognition. Fig. 2 shows the total number of images are used during training for *pre-training + fine-tuning vs random initialization*.

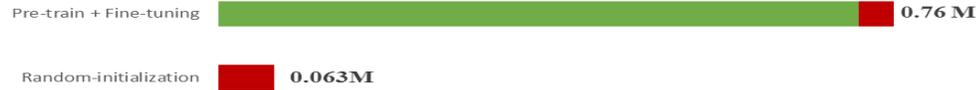

Fig. 2. Total number HD-sEMG images seen during training, for pre-training + fine-tuning vs. random-initialization

Secondly, the network architecture of the existing methods for HD-sEMG image recognition requires pre-training using a large-scale HD-sEMG dataset, therefore the question arises which components of CNNs are necessary for achieving competitive performance as per with these existing methods from *random initialization*. Motivated by the work in [23], we take a first step towards answering this question by studying the simplest architecture we could conceive: a network with consisting of convolution layers, with a maximum of one fully connected layers with a small number of neurons and an occasional dimensionality reduction by using a *max/average pooling* or using a stridden CNN. The use of a small number of convolutions and fully connected layers in S-ConvNet greatly reduce the number of parameters and thus also serve as a form of regularization. The S-ConvNet architectural design are adopted based on the following principles and observations: (i) it is hypothesized that the different muscular activities produce different intensity distributions, which is reproducible across the trials of the same muscular activities and discriminative among different activities. However, we observed that the spatial intensity distributions within the same muscular activities are not locally invariant and the precise location of the features are also independent to the class labels. Fig. 3 demonstrate sequence of HD-sEMG images derived from the same class. The CNN alone have the great abilities to exploit locally translational invariance features through adopting local connectivity and weight sharing strategies [19]. Hence, the LCN are ablated in designing our S-ConvNet models as the location of the features are not dependent to the class labels. (ii) in designing S-ConvNet, we also make use of the fact that if the instantaneous HD-sEMG image is covered by the units in the topmost convolutions layers could be large enough to recognize its content [23].

Thirdly, HD-sEMG image classifier requires *normalization* to help optimization. In addition to deploying successful forms of normalized parameter initialization method [23], [24] employing an effective activation normalization method is equally important when an instantaneous sEMG image recognition model such as S-ConvNet are required to be trained from scratch. Batch Normalization (BN) [25] is a widely used activation normalization technique in the development of deep learning-based methods. BN is used to normalize features by computing the mean and variances over a mini-batches of instantaneous HD-sEMG image samples, which have also shown to be promising in many other applications to ease optimization and enable deep networks to converge faster. Moreover, the stochastic uncertainty of the batch statistics provides some form of regularization which may yield better generalization [26]. In addition to BN, Dropout [27] is another most popular regularization technique and a simple way to prevent deep neural networks from overfitting. However, Dropout and BN often lead to worse performance when they are combined. This is due to the fact that the Dropout would shift the variance of a specific neural unit when the state of the network transfer from training to test. On the other hand, BN would maintain its statistical variance, which is accumulated from the entire training process, in the test phase. This inconsistencies in variance causes the unstable numerical behavior when the signal goes deeper through the network, which may even lead to incorrect predictions [28]. Unlike the existing approaches, the Dropout and BN applied separately in all our S-ConvNet models and evaluated their respective performance.

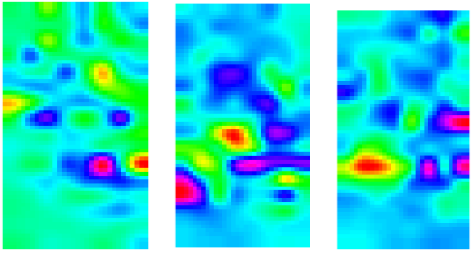

Fig. 3. HD-sEMGs derived from the same muscular activity class which demonstrates that the distributions are independent to the class labels

### A. S-ConvNet Architecture and Training

We train our S-ConvNet on a multi-class neuromuscular activity recognition task, namely, to recognize an activity class through an instantaneous HD-sEMG image. The overall architecture of S-ConvNet models are described in Table I. Starting from the simplest model A, the depth and number of parameters in the network gradually increases to model C. The instantaneous HD-sEMG image is passed through a convolutional (conv.) layers, where a small receptive field with a 3×3 filters are used. The smallest receptive field with 3×3 filters is the minimum filter size to allow overlapping convolutions and spatial-pooling with a stride of 2, which also capture the notion of left, right and center amicably. It can be observed that the model B from the Table I is a variant of the Network in Network architecture [29], where only 1×1 convolution is performed after each normal 3×3 convolutions layers. The 1×1 convolution act as a linear transformation of the input channels followed by a non-linearity [30]. We also highlight that the model C is a variant of the simple ConvNet models introduced by Springenberg *et al.,* [23] for object recognition in which the spatial-pooling is performed by using a stridden CNN. The operation of a convolution map and a subsequent spatial pooling are illustrated in Fig. 4. The output of a convolution map $f$ produced by a convolution layer $c$ is computed as follows:

$$c_{i,j,o}(f) = \emptyset \left( \sum_{h=1}^{k} \sum_{w=1}^{k} \sum_{u=1}^{n} \theta_{h,w,u,o} \cdot f_{g(h,w,i,j,u)} \right) \quad (1)$$

where $\theta$ are the convolutional weights or filters and $g(h, w, i, j, u) = (r.i + h, r.j + w, u)$ is the function mapping from position in $c$ to position in $f$ respecting the stride $r$. $w$ and $h$ are the width and height of the filters and $n$ is the number of channels (in case $f$ is the output of a convolutional layer, $n$ is the number of filters). $o \in [1, M]$ is the number of output feature or channels of the convolutional layer. $\emptyset(\cdot)$ is the activation funtion, an exponential linear unit ELU

TABLE I
THE THREE S-CONVNET NETWORKS MODELS FOR NEUROMUSCULAR ACTIVITY RECOGNITION

| A | B | C |
|---|---|---|
| Input 16×8 Gray-level Image | | |
| 3 × 3 Conv. 32 ELU | 3 × 3 Conv. 32 ELU | 3 × 3 Conv. 32 ELU |
| 3 × 3 Conv. 64 ELU | 1 × 1 Conv. 32 ELU | 3 × 3 Conv. 32 ELU |
| 3 × 3 Conv. 64 ELU | 3 × 3 Conv. 64 ELU | 3 × 3 Conv. 32 ELU with stride $r = 2$ |
| FC1 256 ELU | 1× 1 Conv. 64 ELU | 3 × 3 Conv. 64 ELU |
| FC2 G-way softmax | FC1 256 ELU | 3 × 3 Conv. 64 ELU |
| - | FC2 G-way softmax | 3 × 3 Conv. 64 ELU with stride $r = 2$ |
| - | - | FC1 256 ELU |
| - | - | FC2 G-way softmax |

defined as:
$$\emptyset(x) = \begin{cases} \alpha(\exp(x) - 1), & if \ x < 0 \\ x, & if \ x \geq 0 \end{cases} \quad (2)$$

Then, the spatial pooling operations are performed to the convolution map $f$ as follows:

$$s_{i,j,u}(f) = \left( \sum_{h=1}^{k} \sum_{w=1}^{k} \left| f_{g(h,w,i,j,u)} \right|^p \right)^{\frac{1}{p}} \quad (3)$$

The equation (3) can be interpreted as a *max-pooling* if the $p$ set as $p \to \infty$. In addition, tweaking a little bit of the equation (2) and setting $p = 1$, the following equation can be translated in to an average-*pooling*:

$$s_{i,j,u}(f) = \left( \frac{1}{k} \sum_{h=1}^{k} \sum_{w=1}^{k} \left| f_{g(h,w,i,j,u)} \right|^p \right)^{\frac{1}{p}} \quad (4)$$

The spatial pooling operated on a convolutional map make the networks more robust to local translations and may cope with the electrode shifting problem encountered in different HD-sEMG recording trials and sessions. However, the spatial pooling operations may cause the network to lose information about the detailed texture and micro-textures of an instantaneous sEMG image. Therefore, the pooling operations are only introduced to our models after the first convolutional layers in order to investigate the effect of these pooling operations to our network models.

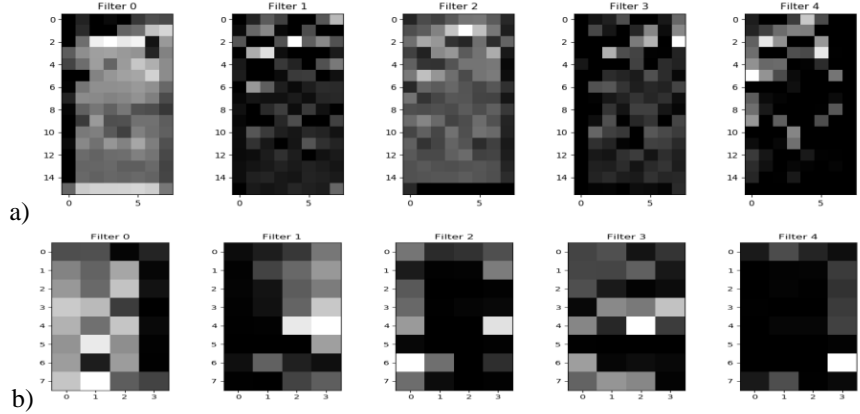

a)

b)

Fig. 4. A schematic illustration of convolutions and pooling operation a) Convolution maps and b) Convolutions maps after spatial pooling.

Afterwards, the convolution maps produced by the final convolutional layer of each of the model networks illustrated in Table I, are flattened out and concatenated to form a multi-dimensional feature vector. Then, the flattened feature vector is inputted to a fully connected layer where each of the feature elements are connected to all its input neurons. This fully connected layer can capture correlations between feature extracted in the distant part of the instantaneous sEMG images. The output of the fully connected layer in the network are used as discriminative feature representation for instantaneous sEMG images. In terms of representation, this is in contrasts to our HOG-based sEMG image representation [16], that generally extract very local descriptors by computing the histograms of oriented gradients and use as input to a classifier.

Finally, the output of the fully connected layer is fed to a G-way softmax layer (where G is the number of neuromuscular activity classes) which produces a distribution over the class labels. If we denote $\hat{y}^{(j)}$ as the $j$th element of the $G$ dimensional output vector of the layer preceding the softmax layer, the class probabilities are estimated using the softmax function $\sigma(.)$ defined as below:

$$\sigma(\hat{y}^{(j)}) = \frac{\exp(\hat{y}^{(j)})}{\sum_G \exp(\hat{y}^{(G)})} \quad (5)$$

The goal of this training is to maximize the probability of the correct neuromuscular activity class. We achieve this by minimizing the cross-entropy loss [31] for each training sample. If $y$ is the true label for a given input, the loss is

$$L = -\sum_j y^{(j)} \ln(\sigma(\hat{y}^{(j)})) \quad (6)$$

The loss is minimized over the parameters by computing the gradient of $L$ with respect to the parameters and by updating the parameters using the state-of-the-art Adam (adaptive moment estimation) gradient descent-based optimization algorithm [32].

Having trained the network, an instantaneous HD-sEMG image is recognized as in the neuromuscular activity class $\hat{g}$ by simply propagating the input image forward and computing: $\hat{g} = argmax_j(\hat{y}^{(j)})$.

B. *Normalization*

As discussed above, the acquired HD-sEMG signals at each sampling instant were arranged in a 2-D grid according to their electrode positioning and converted into an instantaneous sEMG image by linearly transforming the values of sEMG signals from $mV$ to color intensity as $[-2.5mV, 2.5mV]$ to $[0\ 255]$. Therefore, the intensity distribution of the transformed sEMG images are normalized to be between zero and one in order to reduce the sensitivity to contrast and illumination changes. Given an input sEMG image $I$, this is accomplished by applying *max-min* normalization:

$$I'_N = \frac{(I - I_{min}) \cdot (I'_{max} - I'_{min})}{I_{max} - I_{min}} + I'_{min} \quad (7)$$

where $I_{max}$ and $I_{min}$ are respectively the maximum and minimum pixel intensity of the input image $I$. $I'_{max}$ and $I'_{min}$ are the desired maximum and minimum pixel intensity range for the normalized image $I'_N$. It is worth mentioning that our training data were not pre-normalized when the BN is applied.

## IV. THE ALL CONVOLUTIONAL NEURAL NETWORK (ALL-CONVNET)

We find that even our proposed S-ConvNet models described in the previous section can learn all the necessary invariances required for the recognition of instantaneous HD-sEMG images only using the makeshift database avalable for the *target task*. Following this finding, we further propose a new architecture that consists solely of convolutional layers. To design the proposed All-ConvNet model, we make use of the fact that if the image area covered by units in the topmost convolutional layer covers a portion of the image large enough to recognize its content. Then, the fully connected layers can also be replaced by a simple 1-by-1 convolutions. This leads to predictions of HD-sEMG image classes at different positions which can then simply be averaged over the whole image. This scheme was first described by Lin *et al.,* [29], which further regularizes the network as the 1-by-1 convolution has much less parameters than a fully connected layer. Overall our architecture is thus reduced to consist only of convolutional layers with ELU (Eq. 2) non-linearities and an averaging + softmax layer to produce predictions over the whole instantanous HD-sEMG image. Table II describes our proposed All-ConvNet architecture

TABLE II
THE ALL-CONVNET NETWORK MODEL FOR NEUROMUSCULAR ACTIVITY RECOGNITION

| All-ConvNet |
|---|
| Input 16×16 Gray-level Image |
| 3 × 3 Conv.64 ELU |
| 3 × 3 Conv.64 ELU |
| 3 × 3 Conv. 64 ELU with stride $r=2$ |
| 3× 3 Conv. 128 ELU |
| 3× 3 Conv. 128 ELU |
| 3× 3 Conv. 128 ELU with stride $r=2$ |
| 1×1 Conv. 128 ELU |
| 8×8 Conv. 8 ELU |
| global averaging over 8×8 spatial dimensions |
| G-way softmax |

We evaluate the proposed S-ConvNet and All-ConvNet models online by learning the instantaneous sEMG image representation on CapgMyo database for neuromuscular based gesture recognition. Next section discusses experimental results and analysis to evaluate the performance of the proposed S-ConvNet and All-ConvNet as well as some insight and findings about learning and recognizing instantaneous sEMG images.

## V. THE PERFORMANCE EVALUATION OF THE PROPOSED S-CONVNET AND ALL-CONVNET NETWORK MODELS

From the viewpoint of Muscle Computer Interfaces (MCI) application scenarios, the neuromuscular activity recognition can be categorized into two types: I) *intra-session*, in which a classifier is trained on part of the data recorded from the subjects during one session and evaluated on another part of the data recorded from the same session, and II) *inter-session*, in which a classifier is trained on the data recorded from the subjects in one sessions and tested on the data recorded in another session. However, the sEMG-based gesture recognition in the literature has usually been investigated in *intra-session* scenarios [17]. In this preliminary experiments, the performance of our proposed S-ConvNet and All-ConvNet models are evaluated in *intra-session* scenarios. Nevertheless, the implications of our proposed methods in inter-session scenarios will also be reported in our future works.

We evaluated our approach using CapgMyo data sets [17] (this database is made available from following website http://zju-capg.org/myo/data/index.html). This dataset was developed for providing a standard benchmark database (DB) to explore new possibilities for studying next-generation muscle-computer interfaces (MCIs). Table III illustrates gesture in DB-a and DB-b. The CapgMyo database comprises 3 sub-databases (referred as DB-a, DB-b and DB-c). However, DB-a has been used in our preliminary experiments to evaluate the performance of our proposed methods for *intra-session* neuromuscular activity recognition becasue the maximum number of subjects (18) have been participated in DB-a. In DB-a, 8 isotonic and isometric hand gestures were obtained from 18 of the 23 subjects and each gesture was also recorded for 10 times. For each subject, the recorded HD-sEMG data is filtered, sampled and instantaneous sEMG image is generated using the method mentioned in Section II. More explicitly, for each subject 8 different hand gestures are performed and each hand gestures are recorded for 10 times with a $1000_{HZ}$ sampling rate, which in total generates $(8 \times 10 \times 1000 = 80000)$ instantaneous sEMG images individually.

## A. Data Selection for Training, Validation and Testing

Existing approaches for instantaneous HD-sEMG image recognition used a pre-trained model. In

TABLE III.

GESTURES IN DB-A AND DB-B (8 ISOTONIC AND ISOMETRIC HAND CONFIGURATIONS) [17]

| Label | Description | Gesture | Label | Description | Gesture |
|---|---|---|---|---|---|
| 1 | Thumb up | 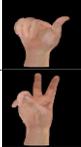 | 5 | Abduction of all fingers | 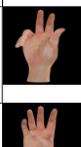 |
| 2 | Extension of index and middle flexion of others | 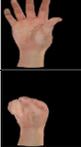 | 6 | Finger flexed together in fist | 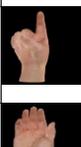 |
| 3 | Flexion of ring and little finger, extension of the others | 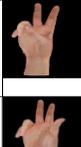 | 7 | Pointing index | 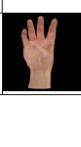 |
| 4 | Thumb opposing base of little finger | 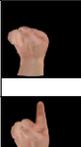 | 8 | Adduction of extended fingers | 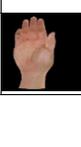 |

pre-training, a total of (18×40000) = 720000 or 0.72 million training images have been used. During re-training (or fine-tuning), 40000 trainings samples for every subject have been used separately. Therefore, the existing approaches involves a total of (720000+40000) = 760000 or 0.76 million images only in the training process. In contrast, our model is learned from scratch through random-initialization. We performed training, validation and testing using only the makeshift datasets available i.e 80000 images produced by 18 subjects individually. In CapgMyo dataset, 8 different hand gestures (Table III) are performed by the 18 different subjects. Also, each of the 8 different hand gestures were repeated 10 times (10 trials) by the 18 different subjects. In every trial, we obtain 1000 images (because the HD-sEMG signals were sampled at 1000Hz). Therefore, we obtain a total of (1000×10×8) = 80000 images individually for each of the 18 different subjects. In order to be able to use a maximum number of images during training, we also introduced and performed a leave one trial out cross-validation. In leave one trial out cross-validation, we kept one trial out from each of the 8 different hand gestures i.e 8000 images for testing. The remaining 9 trials for 8 different hand gestures i.e 72k images have been used for training. From out of this 72k training images, we kept 9k images out randomly for validation to check whether our devised model is overfitting during the training process. Therefore, our training process involves only 63k images which is in contrast to the existing approaches where 760k images are involved in the training process.

Finally, the leave one trial out cross-validation accuracy for 10 different trials are averaged and used as a performance indicator for an intra-subject test.

The cross-validation accuracy A is computed for each class i, as the number of totals correctly recognized hand gestures, c, divided by the total number of tests sEMG images

$$\text{Accuracy, A} = \frac{c}{N} = \frac{\sum c_i}{N} \tag{8}$$

where $i = \{1, 2, ..., G\}$ and G is the number of gesture classes.

For further illustration, the confusion matrix generated from the predicted classification results for leave one trial out cross-validation for subject 2 in the CapgMyo DB-a database are presented in Appendix A.

## B. Experimental Results

In our experiments, we compared all the proposed S-ConvNet models described in Section III on the CapgMyo DB-a dataset without any pre-training or data augmentations. For effective and faster training of a CNN network model with high-dimensional parameter spaces requires a good initialisation strategy for the connection weights, a good activation function, using Batch Normalization and a good regularization technique. The weight initialization scheme, an activation function and the effectiveness of BN and Dropout regularizer are determined in a preliminary experiment using S-ConvNet model A and Subjects 2 from DB-a and then fixed for all other experiments.

We investigate and experiments with two different initialization schemes for the connection weights in Xavier and He initialization [24], [33]. However, we found that the models with He initialization scheme perform on average 1-1.5% worse than the Xavier initialisation counterparts. We hence do not report the results here with the He intitialization to avoid cluttering the experiments. In order to investigate a most suitable activation function for our proposed S-ConvNet models, we also performed experiments with the different activation functions [34], [35]. The results are reported in Table IV. In addition, as the BN and Dropout often lead to worse performance when they are combined as discussed in Section III. In order to investigate this claim, we performed experiments with both of these methods combined and separately. The results are reported in Table IV.

Also, training a CNN network with a high-dimensional parameter spaces requires an efficient optimization algorithm. Objective functions are often stochastic becasue of internal data sub-sampling, dropout regularization and other sources of noise. Hence, we propose to use a computationally efficient stochastic optimization algorithm, Adam [32], which requires only first-order gradients with little memory requirement, is invariant to diagonal rescaling of the gradients and is suitable for high-dimensional problems. It also provides fast and reliable learning convergence that can be considerably faster than the stochastic gradient descent (SGD) optimization algorithm used in the exisitng approaches for instantaneous HD-sEMG image recognition. Our proposed all S-ConvNet and All-ConvNet were trained using Adam optimization algorithms with a momentum decay and scaling decay are intialized to 0.9 and 0.999 respectively. In contrast to SGD, Adam is an adaptive learning rate algorithm, therefore, it requires less tuning of the learning rate hyperparameter. The learning rate 0.001 is intialized to all our experiments. The smaller batches of 256 randomly chosen samples from training dataset are fed to the network during consecutive learning iterations for all our experiments. We set a maximum 100 epochs for training our S-ConvNet and All-ConvNet networks models. However, to avoid overfitting we have also applied early stopping in which the training process are interrupted if no improvments in validation loss are noticed for 5 consecutive epochs. The BN is applied after the input and before each non-linearity. Dropout was used to regularise all networks. The Dropout was applied on all layers with probabilities 35% and 25% for all S-ConvNet and All-ConvNet models respectively. The results for S-ConvNet models A that we conducted for the subject 2 in CapgMyo DB-a database are presented in Table IV.

TABLE IV

NEUROMUSCULAR ACTIVITY RECOGNITION ACCURACY (%) FOR DIFFERENT ACTIVATION FUNCTIONS AND SPATIAL POOLING

| Network | Relu | Leaky-Relu | Elu | Sigmoid | Max-pool | Avg-pool | Avg-run time (min.) |
|---|---|---|---|---|---|---|---|
| A | 95.18 | 95.56 | 93.98 | **95.76** | 94.55 | 94.31 | 2.55 |
| A with BN | 96.16 | 97.34 | 97.50 | **98.00** | 96.66 | 97.13 | 7.74 |
| A with Dropout regularization | **96.99** | 96.68 | 96.58 | 96.30 | 95.19 | 95.61 | 11.27 |
| A with BN and Dropout | 97.18 | 97.54 | **98.29** | 97.80 | 96.98 | 97.54 | 14 |

Several trends can be observed from these results. First, confirming previous results from the literature, the simplest model A (S-ConvNet A) perform remarkably well, achieving 98.29% correct neuromuscular activity recognition rate for an intra-subject test. Second, simply applying *max-min* normalization to the training dataset and fed to the S-ConvNet model A, the average correct recognition rate 94.89% has been achieved for 6 different experiments with only 2.55 min overall runtime for training, validation and testing on a Nvidia Tesla K-20C GPU. Third, simply replacing the *max-min* normalization by introducing BN to the network the average correct recognition rate improved to 97.13% by sparing overall 7.74 min runtime for training, validation and testing. Fourth, the correct recognition rate slightly decreases to 96.23% when the BN is replaced by the Dropout regularizer and *max-min* normalization while also increasing the overall runtime to 11.27 min for entire training, validation and testing process. Fifth, when BN and Dropout with a tiny probability are respectively applied to all layers of the network, the average recognition rate increases up to 97.56%. Due to introducing BN and Dropout to the networks, the overall runtime also increases to about 14 min for entire training, validation and testing process. In all cases, the performance of the model slightly decreases with spatial *max-pooling* and *average-pooling*. However, the spatial pooling can help to regularize CNNs and might be more effective while we will conduct experiments in *inter-session* scenarios. It is worth noting that, the *average-pooling* perform quite well when the BN or BN and Dropout are introduced to the network model (e.g. Table III). All these preliminary experimental results for an intra-subject test confirm that our proposed S-ConvNet models can learn all the necessary invarinaces that requires to build a distinctive representation for instantaneous HD-sEMG image recognition.

From Table IV it can be noticed that the maximum correct recognition rate 98.29% is achieved when *exponential linear unit (ELU)* activation function, BN and a tiny dropout probability are introduced to every layer of the network. Therefore, this configuration is fixed for other S-ConvNet and All-ConvNet models and performed experiments on all 18 subjects participated in CapgMyo DB-a. The results for all S-ConvNet and All-ConvNet models are shown in Table V.

TABLE V

THE AVERAGE RECOGNITION ACCURACY (%) OF 8 HAND GESTURES WITH INSTANTANEOUS HD-SEMG IMAGES FOR 18 DIFFERENT SUBJECTS AND RECOGNITION APPROACHES

| Model | Average Recognition Accuracy (%) | #Learning Parameters (millions) |
|---|---|---|
| **S-ConvNet-A** | **87.69** | **≈ 2.09 M** |
| S-ConvNet-B | 86.94 | ≈ 2.12 M |
| S-ConvNet-C with stride = 2 | 83.92 | ≈ 0.14 M |
| S-ConvNet-C with stride = 1 | 87.02 | ≈ 2.10 M |
| **All-ConvNet** | **85.73** | **≈ 0.008 M** |
| Geng *et.al.,* [13] | 89.3 | ≈ 5.58 M + Pre-training |

As can be seen in the Table V, the simple S-ConvNet models (on the order of 2M learning parameters) trained from *random-initialization* with 3×3 convolutions and a dense layer with only a smaller number of neuron performs comparable to the state of the art for CapgMyo DB-a dataset even though the state of the art methods use more complicated network architectures and training schemes which requires to learn over ≈ 5.58 M parameters during fine-tuning only and also pre-trained with over .72M instantaneous HD-sEMG images. Perhaps even more interesting, the proposed All-ConvNet (on the order of ≈ 0.008 M learning parameters) achieves 85.73% average recognition accuracy of an 8 hand gestures for 18 different subjects. This result is achieved when we performed dimensionality reduction in the All-ConvNet network. As also can be seen in the Table V, the average recognition accuracy decreases when we performed dimensionality reduction in to the network (e.g., row 3). This is due to the fact that the resolution of our instantaneous HD-sEMG images are already very low. Therefore, the recognition accuracy are expected to improve if the proposed All-ConvNet model are trained without performing dimensionality reduction. We are currently re-training the All-ConvNet network on DB-a without performing dimensionality reduction, and will include the results in Table V once training is finished.

Fig. 5 presents the recognition accuracy obtained by our proposed different S-ConvNet and All-ConvNet models for 18 different subjects. We achieve 87.69%, 86.94%, 87.02% and 85.72% average recognition accuracy for the proposed S-ConvNet-A, B, C and All-ConvNet models respectively, which is very compétitive to the more complex, highly resource-bassed and fine-tuned pre-trained models proposed by the existing approaches while also reducing the learning parameters to a large extent. These high recognition accuracies for neuromuscular activity recognition based on instantaneous HD-sEMG image recognition indicate the stability and potentiality of the proposed S-ConvNet and All-ConvNet models. For the fair comparison with the state of the art, the following points are required to be highlighted.

We introduce a leave one trial out cross-validation in which our proposed S-ConvNet and All-ConvNet models are tested with 80000 different samples for every subject. Existing instantaneous HD-sEMG image recognition approaches are tested with 40000 samples for each of the subject. Whereas we have used 80000 samples (twice the number of testing samples) for recognition and achieved the comparable performance on par with the state of the art. It is also noteworthy that the recognition results of all S-ConvNet and All-ConvNet models are obtained without any hyper-parameter tuning. Therefore, we also want to stress out that the results of all models evaluated in this report could potentially be improved or even surpass the state of the art by a thorough hyperparameter tuning.

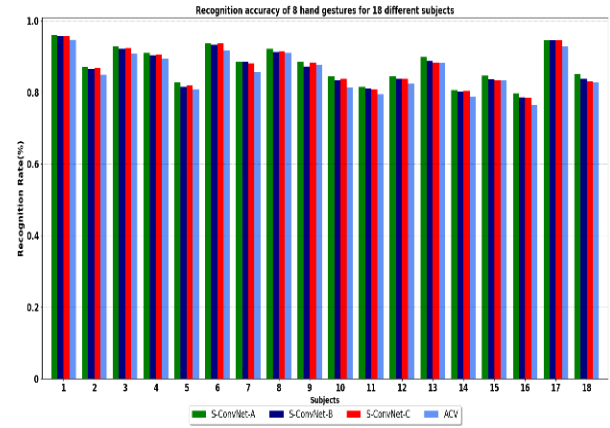

Fig. 5. Recognition accuracy of 8 hands gestures for 18 different subjects with our proposed S-ConvNet and All-ConvNet recognition approaches

Finally, the experimental results demonstrate that,

(i) The proposed S-ConvNet models trained from *random-intialization* can learn all the necessary invariances that requires to build a discriminant representation using only the available target dataset for neuromuscular activity recognition based on instantaneous HD-sEMG images. Therefore, our discoveries will encouage community to devise shallow ConvNet architectures and train the model from the scratch (instead of *pre-training*) for improving the neuromuscular activity recognition performance especially in the data contrained scenarios.

(ii) We definitely agree that, given limitless of training data and unlimited computational power, deep neural networks should perform extremely well. However, our proposed approach and experimental results imply an alternative view to handle this problem : a better S-ConvNet model structure might enable similar or even better performance compared with the more complex existing models trained from large datasets by conducting an exhaustive hyperparameter search. Particularly, our S-ConvNet and All-ConvNet models are only trained with 63K instantaneous HD-sEMG images, but achieve competitve performance as par with the more complex existing models trained with 720k + 40K instantaneous HD-sEMG images. Moreover, as the datasets grow larger, training complex deep neural networks becomes more expensive. Hence, a simple yet efficient approach becomes increasingly significant. Despite its conceptual simplicity, our proposed methods shows a great potential under this settings.

(iii) We argue that, as aforementioned briefly, training from scratch is of critical importance at least for the following reasons. First, *Domain mismatch-* the distributions of the sEMG signals vary considerably even between recording sessions of the same subject within the same experimental set up. This problem becomes more challenging, where the learned model is used to recognize muscular activities in a new recording session. Though the fine-tuning of the pre-trained model can reduce the gap due to the deformations in a new recording session. But, what an amazing thing if we have a technique that can learn HD-sEMG images from scratch for recognizing neuromuscular activities. Second, the fine-tuned pre-trained model restricts the structure design space for neuromuscular activity recognition. This is very critical for the deployment of deep neural networks models to the resource limited scenarios.

(iv) The existing CNN-based neuromuscular activity recognition methods requires a huge memory space to store the massive parameters. Therefore, the models are usually unsuitable for low-end hand-held devices and embedded electronics. Thanks to the proposed parameter-efficient S-ConvNet and All-ConvNet, our model is much smaller than the most competitive methods for instantaneous HD-sEMG image recognition. For instance, our S-ConvNet-A and All-ConvNet model acheives 87.69% and 85.73% average recognition accuracy with only $\approx$ 2.09 M and $\approx$ 0.008 M parameters respectively, which shows a greater potential for applications on low-end devices.

## VI. SUMMARY

We presented S-ConvNet and All-ConvNet models, a simple yet efficient framework for learning instantaneus HD-sEMG images from scratch for neuromuscular activity recognition. Without using any pre-trained models, our proposed S-ConvNet and All-ConvNet demonstartes very competitive accuracy to the state of the art for neuromuscular activity recognition based on instantaneous HD-sEMG images, while using $\approx$ 12 × smaller dataset and reducing learning parameters to a large extent. The proposed S-ConvNet and All-ConvNet has a great potential for learning and recognizing neuromuscular activities on resource-bounded devices. Our future work will consider improving inter-session neuromuscular activity recogntion performances, as well as learning S-ConvNet and All-ConvNet models to support resource bounded devices.

ACKNOWLEDGMENT – This work was supported in part by the regroupement stratégique en microsystèmes du Québec (ReSMiQ) and the Natural Sciences and Engineering Research Council (NSERC) of Canada.

# APPENDIX

## A. A CROSS-VALIDATION RESULTS FOR THE PROPOSED MODEL S-CONVNET A

The confusion matrix generated from the predicted classification results for leave one trial out cross validation for subject 2 in the CapgMyo Db-a database are presented below as an example. The number of correctly classified neuromuscular activities (or hand gesutres) are listed along the diagonal line of the confusion matrix. The mean accuracy (mA) for leave one trial out cross validation are reported below every confusion matrix. The overall recognition rate for subject 2 in the CapgMyo dataset based on our S-ConvNet A is 96.0362%.

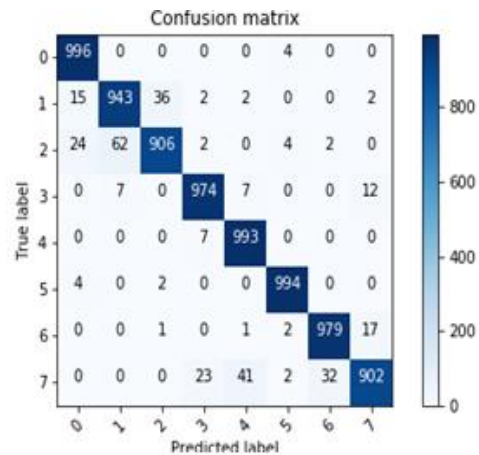

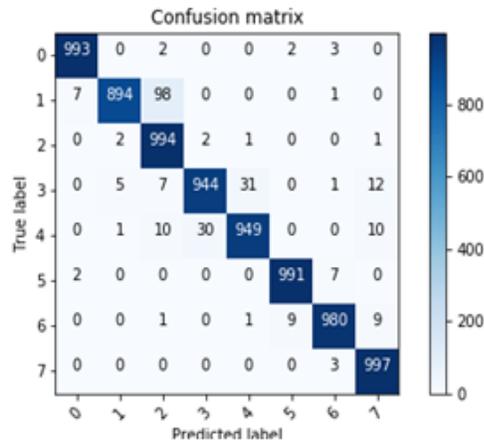

Fig. 6. T1 trial of 8 different hand gestures are used as test set (mA = 96.08)

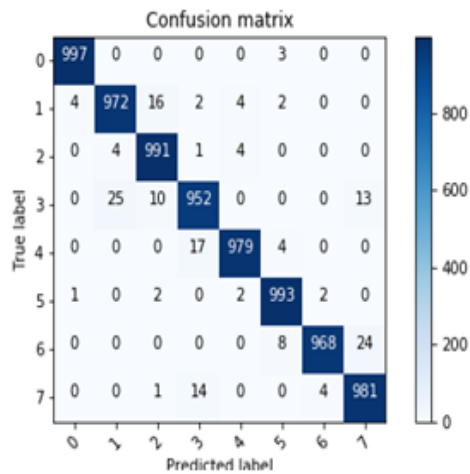

Fig. 7. T2 trial of 8 different hand gestures are used as test set (mA = 96.78)

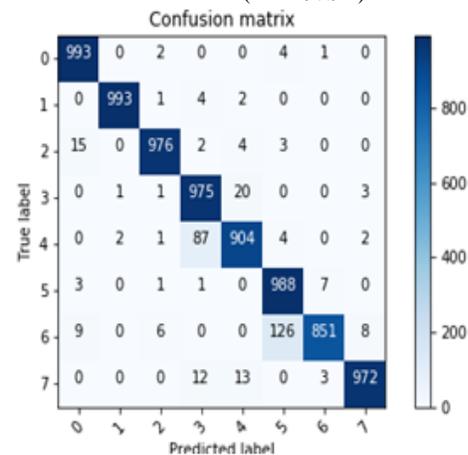

Fig. 8. T3 trial of 8 different hand gestures are used as test set (mA = 97.91)

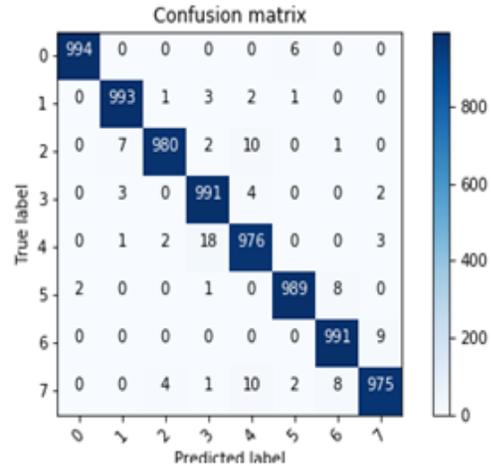

Fig. 9. T4 trial of 8 different hand gestures are used as test set (mA = 95.65)

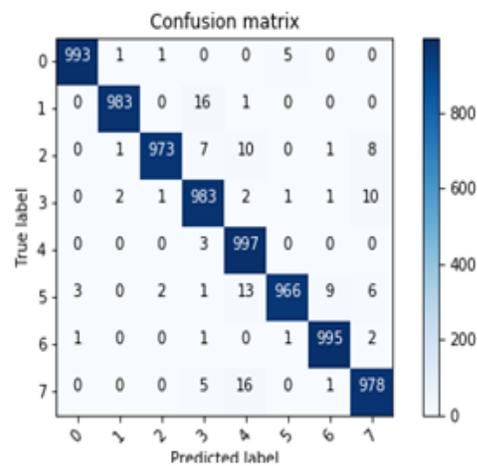

Fig. 10. T5 trial of 8 different hand gestures are used as test set (mA = 98.61)

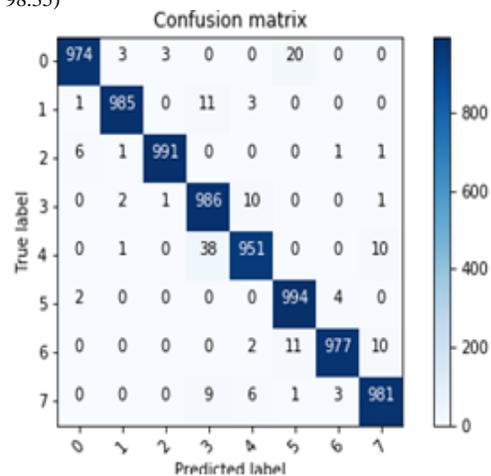

Fig. 11. T6 trial of 8 different hand gestures are used as test set (mA = 98.35)

Fig. 12. T7 trial of 8 different hand gestures are used as test set (mA = 97.98 )

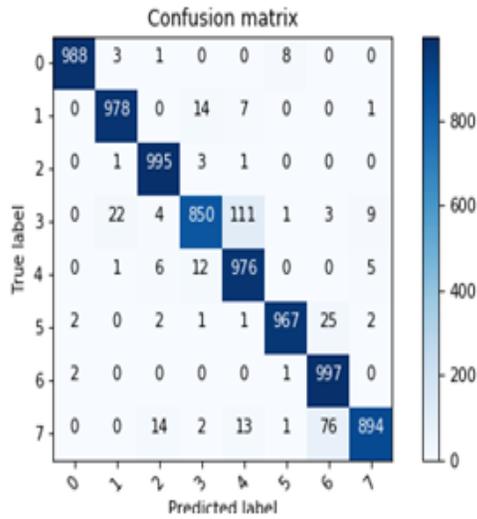

Fig. 14. T9 trial of 8 different hand gestures are used as test set (mA = 94.66)

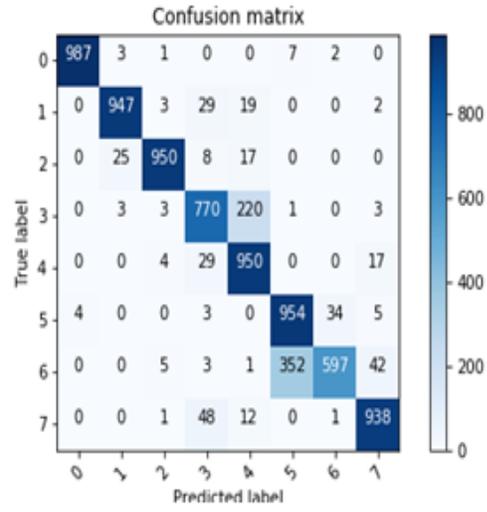

Fig. 13. T8 trial of 8 different hand gestures are used as test set (mA = 95.56 )

Fig. 15. T10 trial of 8 different hand gestures are used as test set (mA = 88.66 )

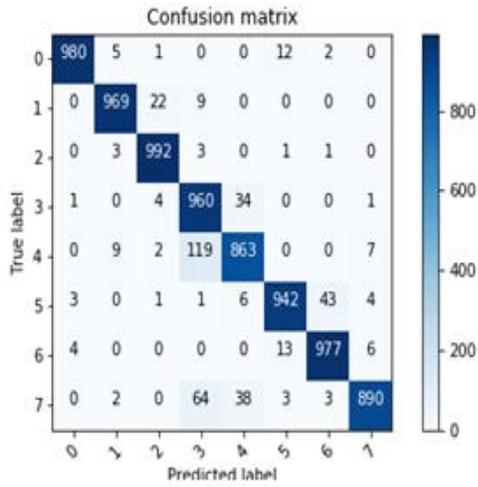